\documentclass[pre,showpacs,showkeys,preprintnumbers,amsmath,amssymb,superscriptaddress,twocolumn]{revtex4-1}
\usepackage[english]{babel}
\usepackage{makeidx} 
\usepackage{graphicx} 
\usepackage{dcolumn} 
\usepackage{array} 
\usepackage{amssymb} 
\usepackage{amsmath}
\usepackage{textcomp}
\usepackage{multirow}
\usepackage{subfigure}
\usepackage{eucal}
\usepackage{mathrsfs}
\usepackage[all]{xy}
\usepackage{epstopdf}

\usepackage{color}

\usepackage{float} 
\usepackage{amsmath} 
\usepackage{amsfonts} 
\usepackage{bm} 

\begin{document}

\title{Characterization of the non--Arrhenius behavior of supercooled liquids by modeling non--additive stochastic systems}


\author{A. C. P. Rosa Jr.}\affiliation{Grupo de Informa\c{c}\~{a}o Qu\^{a}ntica, Centro de Ci\^{e}ncias Exatas e das Tecnologias, Universidade Federal do Oeste da Bahia. Rua Bertioga, 892, Morada Nobre I, 47810-059 Barreiras, Bahia, Brazil.}
\author{C. Cruz\email{clebson.cruz@ufob.edu.br}} \affiliation{Grupo de Informa\c{c}\~{a}o Qu\^{a}ntica, Centro de Ci\^{e}ncias Exatas e das Tecnologias, Universidade Federal do Oeste da Bahia. Rua Bertioga, 892, Morada Nobre I, 47810-059 Barreiras, Bahia, Brazil.}
\author{W. S. Santana}\affiliation{Grupo de Informa\c{c}\~{a}o Qu\^{a}ntica, Centro de Ci\^{e}ncias Exatas e das Tecnologias, Universidade Federal do Oeste da Bahia. Rua Bertioga, 892, Morada Nobre I, 47810-059 Barreiras, Bahia, Brazil.}
\author{M. A. Moret}\affiliation{Programa de Modelagem Computacional - SENAI - CIMATEC, 41650-010 Salvador, Bahia, Brazil}\affiliation{Universidade do Estado da Bahia - UNEB, 41150-000 Salvador, Bahia, Brazil}
\date{\today}

\begin{abstract}

The characterization of the formation mechanisms of amorphous solids is a large avenue for research{,} {since understanding its non-Arrhenius behavior is challenging to overcome}. In this context, we present one path toward modeling the diffusive processes in supercooled liquids near glass transition through a class of non-homogeneous continuity equations, providing a consistent theoretical basis for the physical interpretation of its non-Arrhenius behavior. More precisely, we obtain the generalized drag and diffusion coefficients {that allow us to model} a wide range of non-Arrhenius processes. {{This provides} a reliable measurement of the degree of fragility of the system and an estimation of the fragile--to--strong transition in glass--forming liquids, as well as} a generalized Stokes-Einstein equation, leading to a better understanding of the classical and quantum effects on the dynamics of non--additive stochastic systems.

\end{abstract}
\maketitle

\section{Introduction}

The {dynamic} response of a wide class of materials can be achieved {using} the so-called Arrhenius law \cite{truhlar2001convex,chamberlin1993non,aquilanti2010temperature,kumar2018analysis,rosa2016model}. Basically, it consists of an exponential decay with the inverse of the temperature characterized by the so-called temperature independent activation energy \cite{truhlar2001convex,chamberlin1993non,aquilanti2010temperature}. The search for a physical interpretation for the activation energy established the fundamentals of the transition state theory \cite{truhlar1996current,laidler1983development,meana2011high} since it associates an Arrhenius-like behavior with diffusive processes in several systems \cite{paul2018polymeric,matthiesen2011mixing,dehaoui2015viscosity,kumar2018analysis,frank2015diffusion}. 

However, from the development of new technologies and advances in materials preparation techniques, a wide variety of new compounds could be synthesized,  leading to the improvement of experimental techniques for the study of chemical reactions and diffusive processes. In this scenario, several systems have revealed deviations from Arrhenius behavior, evidenced through the temperature dependence of the activation energy \cite{truhlar2001convex}. 
In {recent} years, the characterization of non-Arrhenius behaviors has received considerable attention, since it was observed in  water type models SPC/E (extended simple point charge) \cite{geske2018molecular,huang2018interplay}, food systems \cite{stroka2011food}, diffusivity in supercooled liquids near glass transition \cite{chamberlin1993non,debenedetti2003supercooled,matthiesen2011mixing,smith2012breaking,dehaoui2015viscosity}, chemical reactions \cite{galamba2016hydrogen,meana2011high,cavalli2014theoretical} and several biological processes \cite{nishiyama2009temperature,roy2017origin}. Therefore, modeling these non-Arrhenius systems {is} a large avenue for research and an actual challenge to overcome. 

The non-Arrhenius behaviors manifest themselves as concave curves (sub-Arrhenius behavior), associated with non-local quantum effects \cite{meana2011high,cavalli2014theoretical,silva2013uniform}, or convex curves  (super-Arrhenius behavior), associated with the predominance of classical transport phenomena \cite{nishiyama2009temperature,aquilanti2010temperature,silva2013uniform,carvalho2017deformed}. 
Despite {much effort} by the scientific community, there are only a few phenomenological relationships proposed to model non-Arrhenius processes, such as the Vogel-Tamman-Fulcher equation \cite{vogel1921h,tammann1926g,fulcher1925analysis} and the Aquilanti-Mundim d-Arrhenius model \cite{agreda2016aquilanti,nishiyama2009temperature,aquilanti2010temperature,cavalli2014theoretical,silva2013uniform,carvalho2017deformed,aquilanti2017kinetics}. {Other} phenomenological expressions have {recently} been proposed \cite{rosa2016model,smith2012breaking,matthiesen2011mixing}{. However, there is a need to establish a wide class of equations that characterize non-Arrhenius processes in a consistent theoretical basis for the physical interpretation of the characteristic  non-Arrhenius behavior of several diffusive processes. }

{Nevertheless}, Aquilanti-Mundim equation can be derived from the stationary {process} of the non-linear Fokker-Planck equation, and the diffusivity dependence with the temperature is consistent with experimental results \cite{rosa2016model}. 
Non-linear Fokker-Planck equations, especially {those} whose stationary solutions maximizes non-additive entropies \cite{schwammle2009dynamics}, such as the Tsallis entropy \cite{tsallis1988c}, has been successfully employed for modeling non-Markovian processes \cite{sicuro2016nonlinear,dos2017random}, anomalous diffusion \cite{marin2018nonlinear,ribeiro2017multi}, astrophysical systems \cite{moret2010x}, sunspots \cite{moret2014self} and pitting corrosion \cite{rosa2015non}, suggesting that this class of equations can also be an alternative way to describe the non-Arrhenius behavior of non-additive stochastic systems.

In this context, we show in this letter a class of non-homogeneous continuity equations whose the generalized coefficient allow {the modeling of} a wide range of non-Arrhenius processes. We modeled the characteristic super-Arrhenius behavior of diffusivity and viscosity in supercooled liquids, {{determining} a characteristic threshold temperature associated with the discontinuities in its dynamic properties, such as the viscosity and the activation energy. In addition, we define a generalized exponent {that} characterizes the non-Arrhenius process and serves as an indicator of the level of fragility in glass-forming systems, whereas the threshold temperature indicates a fragile--to--strong transition, {the} general behavior of metallic glass--forming liquids \cite{Mauro19780}.} {Our} model {also derives} a generalized version for the Stokes-Einstein equation, where we obtain a characteristic temperature independent behavior (at low temperatures) for sub-Arrhenius processes, and a sudden death behavior around the {threshold temperature} for super-Arrhenius processes. 
Our results pave the way {for} the characterization of the breakdown of the standard Stokes-Einstein relation \cite{wei2018breakdown,kohler2017breakdown,ohtori2017breakdown,pan2017structural}, mainly in supercooled liquids \cite{sosso2012breakdown}, providing one path toward understanding the dynamic evolution of non-Arrhenius processes,
leading to the establishment of a theoretical interface between a macroscopic and microscopic {perspective} of the matter through a non-equilibrium statistical mechanics.

\section{Generalized Reaction--Diffusion Model}

Let us consider a concentration $\rho(r,t)$ of a substance measured in volume $V$ at time $t$, the total amount of substance for the same volume is given by the non-homogeneous continuity equation. In this context, we propose the following conditions: 
\begin{itemize}
\item[(i)] $f(r,t)= \vec{\nabla}\cdot \vec{\eta}(r,t)$ is a volumetric density per unit time associated with dissipative processes and $\vec{\eta}(r,t)$ is a field of non-zero divergence;
\item[{(ii)}] $\vec{\eta}(r,t) = -\kappa_{m}^{-1}\rho^{m} \vec{\nabla} \phi$, where $\kappa_{m}$ is a positive constant parameterized by the exponent $m$ and $\phi$ is a generalized potential;
\item[{(iii)}] for the steady state $\vec{\eta}(r,t)\rightarrow \vec{\eta}_S(r)$, {which} is a field of zero divergence; 
\item[{(iv)}] $\vec{J}=-D(r,t;\rho)\vec{\nabla} \rho$ is a diffusion flux, for a generalized version of Fick's first law \cite{frank2005nonlinear,frank2015diffusion} in which $D(r,t;\rho)$ is a generalized diffusion coefficient; 
\item[{(v)}] $D(r,t;\rho)= (\Gamma/2) \rho^{n-1}$ \cite{schwammle2009dynamics}, where $\Gamma$ is a positive definite parameter, related to a class of nonlinear equations associated with anomalous diffusive processes \cite{zanette1995thermodynamics}.
\end{itemize}


In this circunstances, the non-homogeneous continuity equation becomes a particular class of nonlinear Fokker-Planck equations \cite{schwammle2009dynamics} whose non-linearity of the generalized drag coefficient involves the information of the dissipative or exchange processes, such as phase transitions or chemical reactions. 
The equations are defined in such a way that leads to the generation of solutions that compose a class of rapidly decreasing functions \cite{beerends2003fourier} that maximizes non-additive entropies, such as the Tsallis entropy \cite{tsallis1988c}, since this guarantees the possibility of fundamental solutions for the diffusion equation. From these conditions we obtain an alternative way to describe the non-Arrhenius behavior of the diffusion processes of non-additive stochastic systems such as supercooled liquids, from a consistent theoretical basis.

\section{Diffusivity and viscosity of glass--forming liquids}

The characterization of diffusivity and viscosity in supercooled liquids are effective  {to understand the} glass transition and the formation mechanisms of amorphous solids.
In order {to} establish a wide class of equations that characterize non--Arrhenius behavior of supercooled liquids {from} a theoretical {perspective}, we define the diffusion coefficient {in (v)} for the particular case $n=2$. 
In this context, the generalized potential $\phi$ can be reinterpreted as a potential energy $U\left(r\right)$ associated {with} a conservative force field, in dynamic equilibrium. {Thus, we obtain the non-homogeneous continuity equation:}
\begin{equation}
\frac{\partial \rho(r,t)}{\partial t}= \kappa_{m}^{-1}\vec{\nabla} \cdot\left[ \left( \vec{\nabla} U\left(r\right)\right) \rho^{m}\right] + \frac{\Gamma}{2}{\nabla}^2 \left[ \rho^2\right]~.
\label{eq:05}
\end{equation}

{Because} the stationary solution of Eq. (\ref{eq:05}) is a generalized exponential, the dependence of the generalized diffusion coefficient with the temperature {can be} written as,
\begin{equation}
D(T)= D_0 \left[ 1-\left( 2-m\right)\frac{E}{k_BT}\right]^{\frac{1}{2-m}}~,
\label{eq:07}
\end{equation}
where $D_{0}=\Gamma C_{0}$ ($C_0$ is a normalization constant of the stationary concentration), $E=-\int \vec{\nabla}U\left(r\right)\cdot r$ is a generalized energy and $C_0^{2-m}\kappa_{m}\Gamma = k_BT$ \cite{dill2012molecular,islam2004einstein}. 
From Eq. (\ref{eq:07}), the Arrhenius standard behavior is recovered when the coefficient $m\rightarrow 2$, then the activation energy $E$, in this limit, corresponds to a temperature independent energy.

\begin{figure}[htp]
	\centering
	{\includegraphics[scale=0.45]{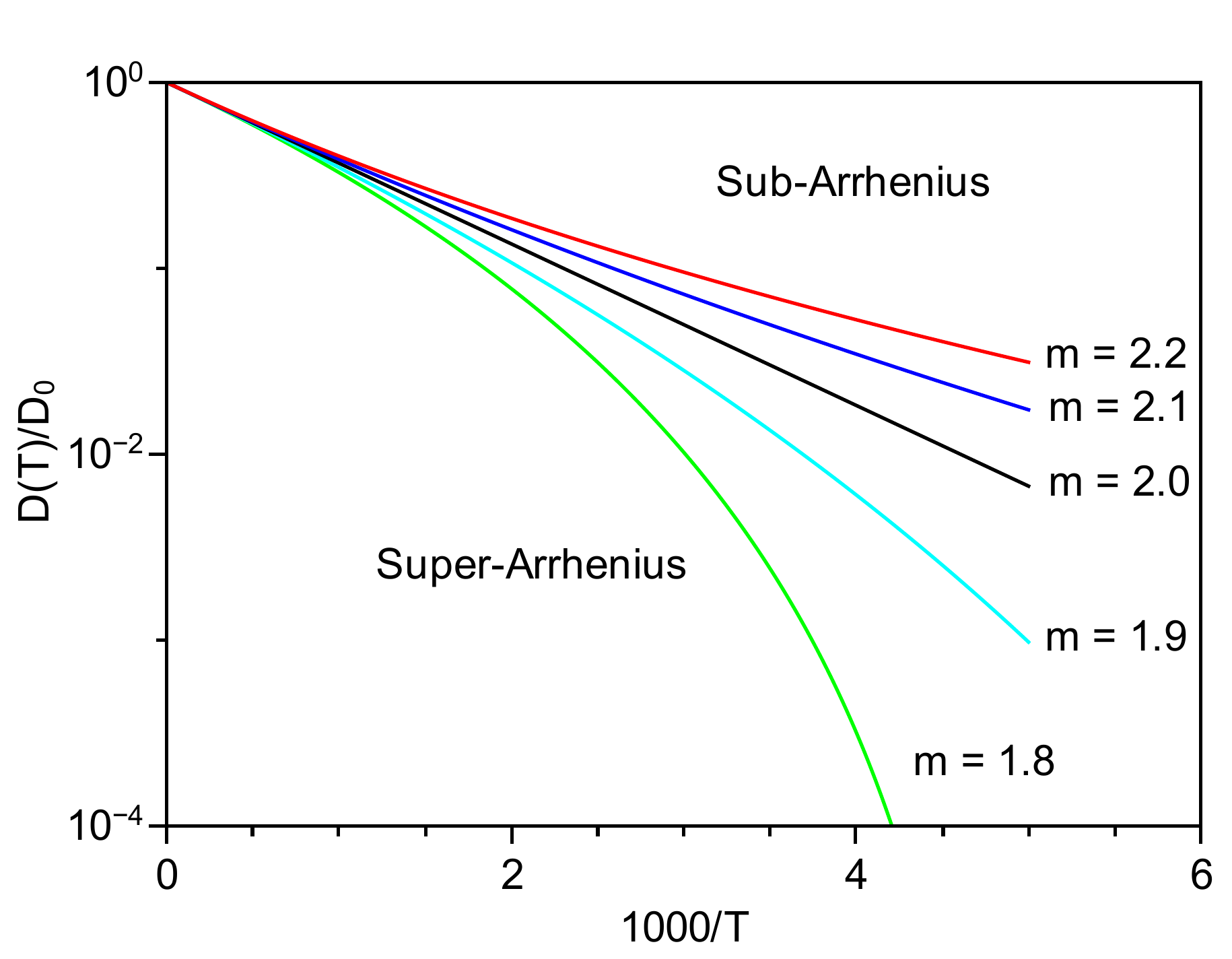}}\\
	\caption{Monolog plot of the diffusivity as a function of {the} reciprocal temperature. The curves $m> 2$ {characterize} a class of sub-Arrhenius processes, while the curves  $m <2$  characterizes a class of super-Arrhenius processes. The $m=2$ curve corresponds to the usual Arrhenius plot. The curves were simulated for {the} $E / k_B = 1000 K$ condition}
	\label{fig:fig1}
\end{figure}

Figure \ref{fig:fig1} shows the diffusivity of a supercooled liquid as {a} function of {the} reciprocal temperature. {{Under} the condition $m <2$ the proposed model encompasses a class of super-Arrhenius diffusive processes, associated with the predominance of classical transport phenomena \cite{nishiyama2009temperature,aquilanti2010temperature,silva2013uniform,carvalho2017deformed}, predominantly according {to experimental} reports \cite{debenedetti2003supercooled,matthiesen2011mixing,smith2012breaking,rosa2016model}. In addition,  the model also covers a wide class of sub-Arrhenius diffusive processes, characterized by the condition $m > 2$, associated with non-local quantum effects \cite{meana2011high,cavalli2014theoretical,silva2013uniform}, and less sensitive to the exponent variations than the super-Arrhenius processes.}

{It} is also possible to verify the existence of a threshold temperature for super-Arrhenius processes, from which the diffusivity goes to zero, given by
\begin{equation}
T_{t}= \frac{\left( 2-m\right)E}{k_B}
\label{eq:tt}
\end{equation}

From Eq. (\ref{eq:07}) we can obtain the temperature dependence of the activation energy  as  
\begin{equation}
E_A\left( T\right)= \frac{E}{1-\left( 2-m\right)\frac{E}{k_BT}} ~,
\label{eq:09}
\end{equation}
the main feature of non--Arrhenius processes. Furthermore, from Eq. (\ref{eq:09}), for the $m\rightarrow 2$ the activation energy achieves a temperature independent behavior $E_A(T)\rightarrow E$ corresponding to the Arrhenius law, as {previously mentioned.}

\begin{figure}[htp]
\centering
{\includegraphics[scale=0.45]{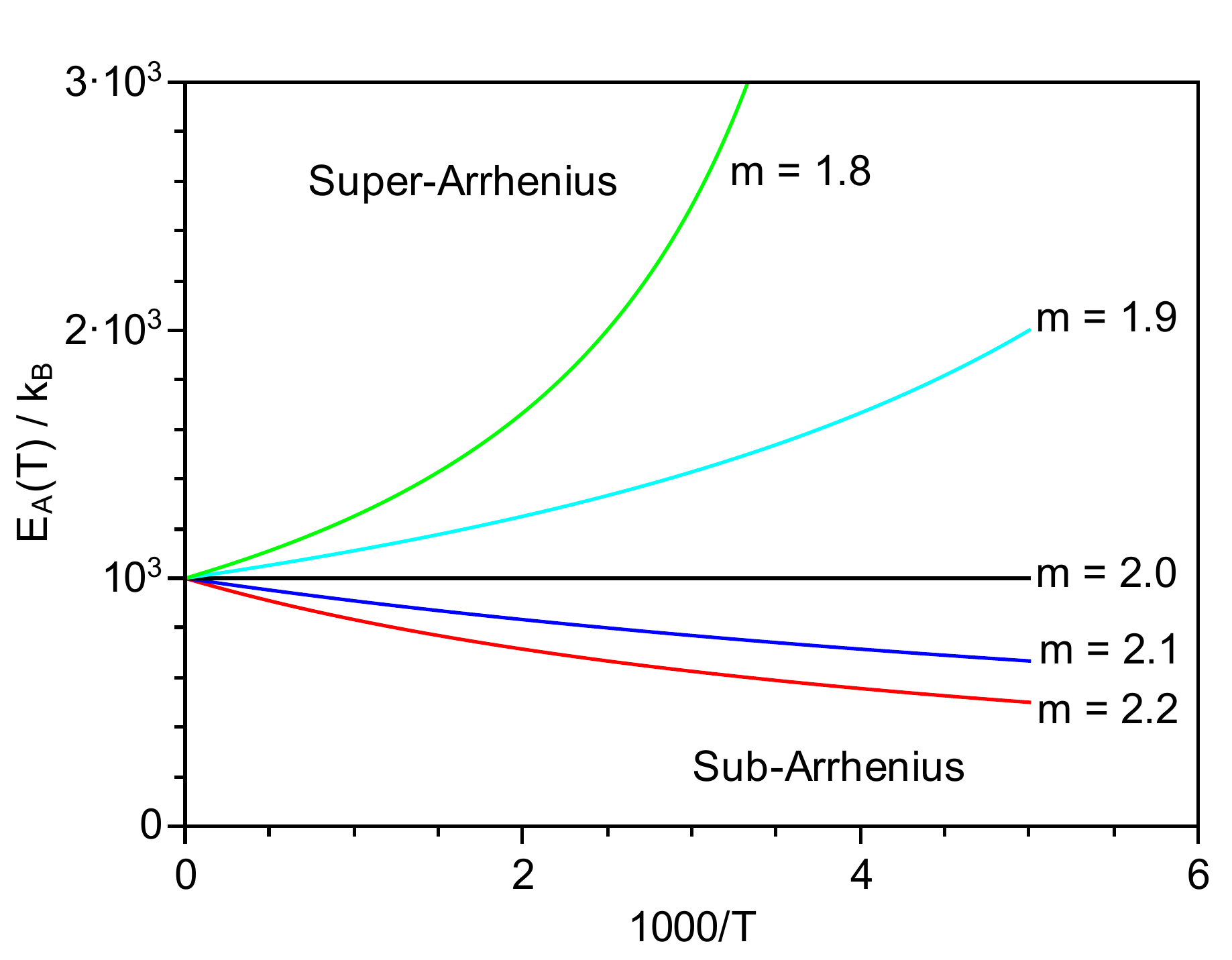}}\\
\caption{{The activation energy as a function} of the reciprocal temperature. The curves where the activation energy is an decreasing function of the reciprocal temperature characterize a class of sub-Arrhenius processes, while the increasing {curves} characterize a class of super-Arrhenius processes. In addition, {the} $m=2$ curve corresponds to the Arrhenius activation energy, characterized by a temperature independent behavior. The curves were simulated using the scale factor $E / kB = 1000 K$.}
\label{fig:fig2}
\end{figure}

Figure \ref{fig:fig2} shows the activation energies, corresponding to the diffusivity curves {presented} in Figure \ref{fig:fig1}, calculated {from} Eq. (\ref{eq:09}). {The} activation energy is an increasing function of the reciprocal temperature for sub-Arrhenius processes and decreasing for super-Arrhenius processes. In addition, for the super-Arrhenius processes, when the threshold temperature{, Eq. (\ref{eq:tt}),} is achieved the activation energy diverges to infinity,  indicating that this temperature {is related} to the viscosity divergence in the glass transition.

{{From} Eq. (\ref{eq:05}), we can define the viscosity from the generalized mobility of the fluid \cite{dill2012molecular}} as 
\begin{equation}
\eta\left(T\right)= \alpha\kappa_{m}\rho^{1-m}~,
\label{eq:10}
\end{equation}
where $\alpha$ is a positive definite constant. From  Eq. (\ref{eq:10}) the Arrhenius model from {the} viscosity is recovered for the limit case  $m\rightarrow 2$. 

\begin{figure}[htp]
	\centering
	{\includegraphics[scale=0.45]{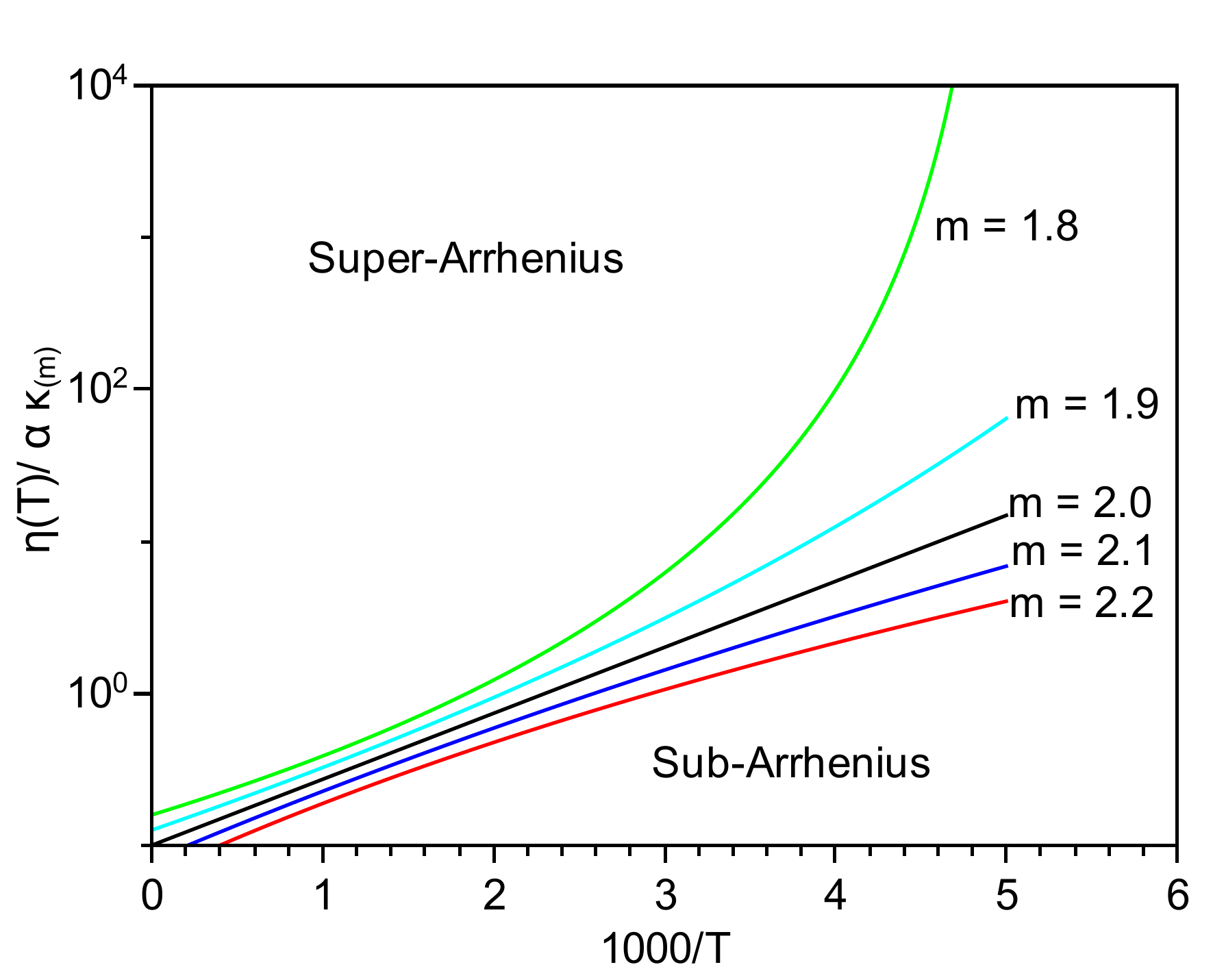}}\\
	\caption{{The viscosity} as a function of the reciprocal temperature. The {curves $m> 2$ characterize a class of sub-Arrhenius models for viscosity, while the curves  $m <2$  characterize} a class of super-Arrhenius models. The $m=2$ curve corresponds to the Arrhenius model for the viscosity.}
	\label{fig:fig3}
\end{figure}

Figure \ref{fig:fig3} shows the viscosity as a function of the reciprocal temperature. For super-Arrhenius processes ($m <2$) the threshold temperature characterizes the regime from which the viscosity diverges to infinity. {{Thus}, the threshold temperature, Eq. (\ref{eq:tt}), serves as an {indication} of how close the system is {to} the glass transition region because it involves discontinuities in {the} dynamic properties, such as the activation energy, Eq. (\ref{eq:09}), and viscosity, Eq. (\ref{eq:10}). The glass-liquid transition occurs {in} a range of temperatures for which {the} viscosity {assumes} a large value, but still does not diverge. In most glass-forming liquids, the glass transition temperature is established {at} the viscosity reference value of $10^{12}$ Pa.s, thus $T_t \leq T_g$.}
 
{This model can also be used to} calculate the level of fragility $M_\eta$ in glass-forming systems \cite{Mauro19780,angell1988perspective,martinez2001thermodynamic} by our exponent $m$ as
\begin{equation}
M_\eta = \left( \frac{m-1}{2-m}\right) \left(\frac{ 1 }{1-\frac{T_t}{T_g}}\right)~,
\label{eq:11}
\end{equation}

{For {the} usual Arrhenius diffusive processes, the condition $m = 2$ characterizes a \textit{strong} glass system, whereas for a wide class of super-Arrhenius diffusive processes the condition $m < 2$ characterizes a \textit{fragile} glass \cite{angell1988perspective,martinez2001thermodynamic}. {In addition, another important feature that arises from our model is the distinguishability between strong and fragile systems for super-Arrhenius processes ($m < 2$), since how far further the glass transition temperature $T_g$ is from the threshold temperature, Eq. (\ref{eq:tt}), more fragile the system will be. In this way, the ratio $T_t/ T_g$ (Eq. \ref{eq:11}) indicates a  \textit{fragile--to--strong transition} \cite{Mauro19780} {usually} found in some water and silica systems, {which is} possibly a general behavior of metallic glass--forming liquids \cite{Mauro19780}, where an initially fragile supercooled liquid can be transformed into a strong liquid upon supercooling toward $T_g$. Therefore, the dynamics around the glass transition region, characterized by Eq. (\ref{eq:tt}) {provide} a measurement of how fragile a system is, establishing the {theoretical basis understanding the intrinsic} features of the formation mechanisms of amorphous solids.}}


Moreover, a {remarkable} result can be extracted from our model. The product between the generalized diffusion coefficient, Eq. (\ref{eq:07}) and the viscosity, Eq. (\ref{eq:10}), obtained from our generalized model for reaction--diffusion processes, provides a generalized Stokes-Einstein relation for any non--Arrhenius diffusion process, given by
\begin{equation}
D \eta = \alpha k_BT  \left[ 1-\left( 2-m\right)\frac{E}{k_BT}\right] 
\label{eq:12}
\end{equation}

\begin{figure}[htp]
	\centering
	{\includegraphics[scale=0.45]{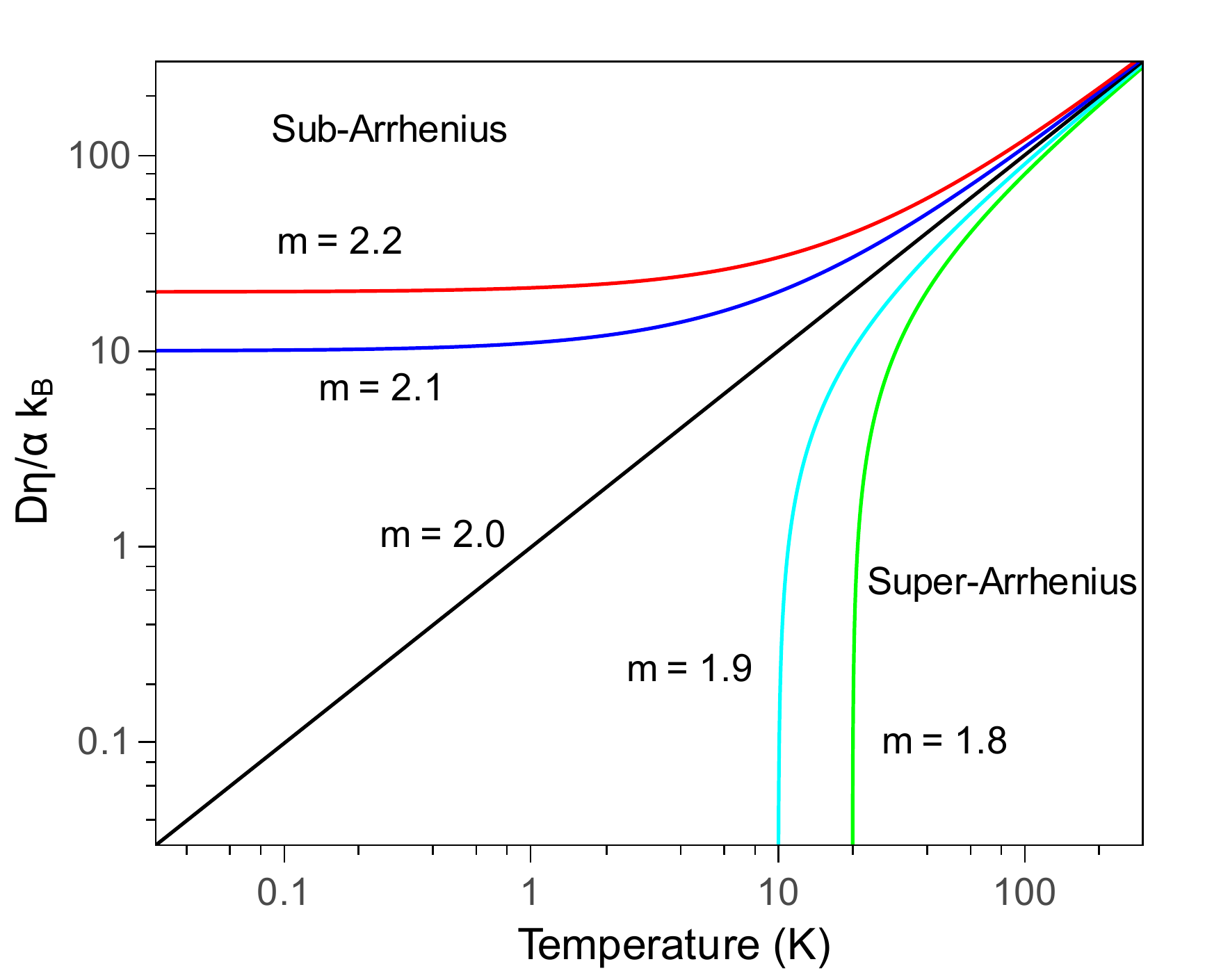}}\\
	\caption{{The temperature} dependence of the generalized Stokes-Einstein relation, Eq. (\ref{eq:12}),  for different values of the coefficient $m$. For super-Arrhenius processes ($m <2$) the region in which the generalized Stokes-Einstein rapidly goes to zero is equivalent to the threshold temperature of glass transition,  Eq. (\ref{eq:tt}). For sub-Arrhenius processes, ($m >2$), from the condition $E >> k_B T$ the generalized Stokes-Einstein equation presents a temperature independent behavior. The straight line corresponds to the  usual form of the Stokes-Einstein relation, recovered for any Arrhenius-like process ($m= 2$) and for the condition $E << k_B T$, that separates the super and sub-Arrhenius regimes.}
	\label{fig:fig4}
\end{figure}

Figure \ref{fig:fig4} shows the temperature dependence of the generalized Stokes-Einstein relation, Eq. (\ref{eq:12}),  for different values of the coefficient $m$. For the super-Arrhenius diffusive processes ($m <2$) the relation gives an estimate of the glass transition temperature, since the generalized diffusion coefficient, Eq. (\ref{eq:07}),  goes to zero faster than the viscosity, Eq.(\ref{eq:11}), diverges to infinity. Thus, the region in which the generalized Stokes-Einstein goes to zero is equivalent to the threshold temperature of glass transition,  Eq. (\ref{eq:tt}).
In addition, as {demonstrated in} Figure \ref{fig:fig4}, the usual form of the Stokes-Einstein relation is recovered from Eq. (\ref{eq:12}) {under} two conditions: (i) for any Arrhenius-like process ($m\rightarrow 2$); and (ii) for the condition $E << k_B T$, i.e., thermal fluctuations predominate in the process, to the detriment of the concentration gradient.

On the other hand, for the sub-Arrhenius diffusive processes, it is worth noting that, from the condition $E >> k_B T${,} the generalized Stokes-Einstein equation, Eq. (\ref{eq:12}), presents a temperature independent behavior, enabling the differentiation of the classical and quantum regimes, paving the way {for} the characterization of sub-Arrhenius processes through Eq. (\ref{eq:12}). {This provides} one path {toward understanding the quantum effects} in the dynamics of the non--additive stochastic systems.

\section{Conclusions}

In summary, {our main result was to provide an alternative way} to describe the non--Arrhenius behavior of diffusive processes in glass-forming liquids. 
{Our model was characterized by a generalized exponent $m$ {that defines} the class of non-Arrhenius processes and serves as an indicator of the degree of the fragility in these systems. In addition, we determine the threshold temperature, Eq. (\ref{eq:tt}), from which the dynamic properties, such as the activation energy and viscosity diverges, and gives us a reliable estimate of the degree of fragility, since the ratio $T_t/ T_g$ (Eq. \ref{eq:11}) indicates a  fragile--to--strong transition, establishing the theoretical basis {for understanding the intrinsic} features of amorphous solids. }

Also {interesting} is the realization of a generalized Stokes-Einstein equation, Eq. (\ref{eq:12}), which allows us to characterize the breakdown of the standard Stokes-Einstein relation in supercooled liquids. For sub-Arrhenius processes, the generalized relation presents a characteristic temperature independent behavior at low temperatures while, for the class of super-Arrhenius diffusive processes, rapidly goes to zero around the threshold temperature. 
Moreover, the usual form of the Stokes-Einstein relation is recovered for any Arrhenius-like process and when the thermal fluctuations predominate in the process to the detriment of the concentration gradient ($E << k_B T$). Our results provide one path toward the differentiation of the super and sub-Arrhenius processes, leading to a better understanding of {the} classical and quantum effects on the dynamics of non--additive stochastic systems, paving the way {for} the characterization of the formation mechanisms of amorphous solids through the study of non-Arrhenius diffusive processes in these systems.

\begin{acknowledgments}
{The authors thank} James C. Phillips for his helpful comments. This study was financed in part by the CNPq and  the \textit{Coordena\c{c}\~{a}o de Aperfei\c{c}oamento de Pessoal de N\'{i}vel Superior - Brasil} (CAPES) - Finance Code 001.
\end{acknowledgments}

\end{document}